\documentclass{article}

\usepackage{jsr-style}

\title{\textbf{Helium-Electrospray: an improved sample delivery system for single-particle imaging with X-ray lasers}}

\date{} 

\author[1]{Tej Varma Yenupuri\(^{\ddag}\)}
\author[2,3]{Safi Rafie-Zinedine\(^{\ddag}\)} 
\author[1]{Lena Worbs}
\author[3]{Michael Heymann}
\author[2]{Joachim Schulz}
\author[2]{Johan Bielecki\(^{*}\)}
\author[1,4]{Filipe R. N. C. Maia\(^{*}\)}

\affil[1]{\textit{Laboratory of Molecular Biophysics, Department of Cell and Molecular Biology, Uppsala University, Husargatan 3 (Box 596), Uppsala, 75124, Sweden}}
\affil[2]{\textit{European XFEL, Holzkoppel 4, 22869 Schenefeld, Germany}}
\affil[3]{\textit{Institute of Biomaterials and Biomolecular Systems, University of Stuttgart, Pfaffenwaldring 57, Stuttgart, 70569, Germany}}
\affil[4]{\textit{Lawrence Berkeley National Laboratory, Berkeley, CA, 94720, USA}}
\begin{document}
\maketitle
\begin{tabular}{@{}l@{}}
\(^{\ddag}\)These authors contributed equally to this work.\\ 
\(^{*}\) Correspondence e-mail: johan.bielecki@xfel.eu, filipe.maia@icm.uu.se
\end{tabular}
\vspace{0.3in}

\begin{abstract} 
Imaging the structure and observing the dynamics of isolated proteins using single-particle X-ray diffractive imaging (SPI) is one of the potential applications of X-ray free-electron lasers (XFELs). Currently, SPI experiments on isolated proteins are limited by three factors: low signal strength, limited data and high background from gas scattering. The last two factors are largely due to the shortcomings of the aerosol sample delivery methods in use. Here we present our modified electrospray ionization (ESI) source, which we dubbed Helium-ESI (He-ESI). With it, we increased particle delivery into the interaction region by a factor of 10, for 26 nm-sized biological particles, and decreased the gas load in the interaction chamber corresponding to an  80\% reduction in gas scattering when compared to the original ESI.
These improvements will lead to a significant increase in the quality and quantity of SPI diffraction patterns in future experiments using He-ESI, resulting in higher-resolution structures.


\end{abstract}


\section{Introduction} 

Current generation X-ray free electron lasers (XFELs) with their ability to produce highly intense X-ray pulses with durations of only a few tens of femtoseconds offer a powerful tool to image a wide variety of aerosolized particles at room temperature. Such high intensities on femtosecond time scales suggested that useful data could be collected from weakly scattering single proteins or viruses by outrunning radiation damage using the idea of “diffraction before destruction” \cite{neutze2000potential}. Taking full advantage of this new capability of coherent diffractive X-ray imaging using single particles in the gas phase promises to not only deliver high-resolution structures but to extend the study towards ultrafast dynamics \cite{Beyond_crystallography, aquila2015linac}, opening the door for pump-probe experiments on femto-and picosecond time scales. 
So far, single-particle imaging (SPI) experiments have been successfully performed by injecting the aerosolized sample into the X-ray interaction region using the "Uppsala"-injector \cite{hantke2018rayleigh} on large biological samples (70–2000 nm) using gas dynamic virtual nozzles (GDVN's) on viruses \cite{seibert2011single, ekeberg2015three, munke2016coherent, reddy2017coherent, lundholm_2018, rose_2018}, cell organelles \cite{hantke2014high}, whole cells \cite{van2015imaging} and most recently on gold nanoparticles \cite{Ayyer:2021gold} using electrospray ionization (ESI). 

\begin{figure}[ht]
    \centering
    \includegraphics[scale=0.12]{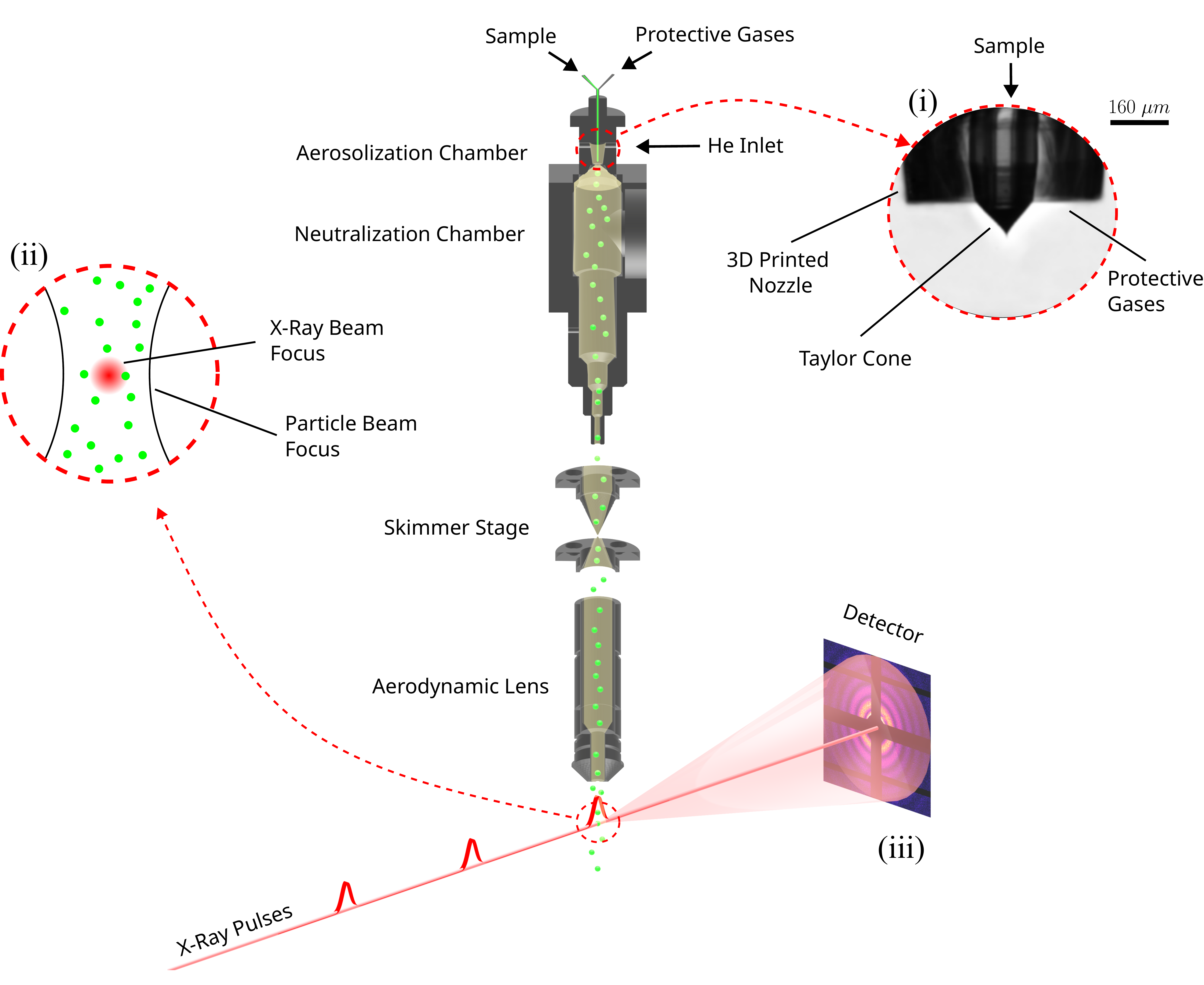}
    \caption{The schematic diagram details a typical experimental setup of an electrospray-based aerosol injector used for single particle imaging experiments at free electron lasers.
    This setup includes the ESI process illustrated at the top, which aerosolizes the sample. Subsequently, the aerosol beam is transported through the skimmer stages and the aerodynamic lens and eventually reaches the interaction chamber. Here, it intersects with the XFEL beam. The XFEL pulses scatter off the particles within the aerosol beam, generating diffraction patterns captured on the detector. (i) The Taylor cone, during standard operation of He-ESI.
    (ii) The interaction between a particle beam and an X-ray beam. (iii) Scattering pattern produced by a particle.} 
    \label{fig:xfel}
\end{figure}

Gas phase injection \cite{bogan2008single, worbs2021optimizing} via an aerodynamic lens stack (ALS) has gained substantial attention for its high scattering contrast, low background scattering compared to liquid sample delivery, capacity for high-rate data collection and wide sample compatibility. The typical experimental SPI layout is shown in Figure \ref{fig:xfel}. 
Particularly, ESI as a sample aerosolization method has proven effective due to its ability to produce small droplets, resulting in virtually contaminant-free sample delivery \cite{electrospray_bielecki_2019}. But even with the large pulse energies available at modern XFEL facilities, the diffraction patterns from small particles, such as single proteins or virus particles with sizes smaller than 50 nm have a very low signal-to-noise ratio preventing structure determination, despite computational efforts to reduce the noise \cite{bellisario2022noise}. A recent experiment on the GroEL complex from {\it E. coli} delivered using ESI highlights the challenge for small bioparticles: a high amount of background scattering from the N$_2$ and CO$_2$ in the interaction region \cite{ekeberg2022observation}. The large gas background hampers the identification of signal from the sample of interest.  

To obtain a higher-resolution structure with current sample injection and XFEL facility parameters, a reduction of N$_2$ and CO$_2$ gas density in the interaction region and a higher particle throughput, i.e., higher hit rates to collect several hundred thousand hits \cite{pandey2020time} and signal averaging over many identical particles are needed. Collecting this amount of data on identical particles makes sample delivery techniques one of the crucial factors in achieving high-resolution atomic-scale images at high acquisition rates \cite{hantke2014high,sobolev2020megahertz}. 
  
In this paper, we address these sample delivery challenges and present a modified ESI source, which we refer to as the Helium electrospray (He-ESI). The main change is the addition of a 3D-printed nozzle, designed to reduce the N$_2$ and CO$_2$ consumption compared to the earlier setup (original ESI) \cite{electrospray_bielecki_2019} while still maintaining stable sample delivery conditions. Helium (He) is introduced around the 3D-printed nozzle and serves as the main gas for particle transport. Our modifications lead to a lower N$_2$ and CO$_2$ use and a decrease of heavy gasses in the interaction region by ~83\%. We also demonstrate the successful use of the He-ESI with the "Uppsala"-injector and compare the performance with the original ESI in the injector setup. We observe an increase in injection yield which can be as high as a factor of ~10 for the small biological particles. 

Our He-ESI system shows great potential for SPI of small particles. The reduction in heavy-gas background effectively increases the signal-to-noise ratio. Furthermore, the use of He as the transport gas improves particle focusing in the "Uppsala"-injector, and enhances the throughput of particles into the interaction region. The ESI-setup developed here makes it possible to acquire millions of diffraction patterns with sufficiently low background, an important milestone on the way to high-resolution time-resolved 3D structures of isolated proteins and viruses using SPI.

\section{Methods and Results}\label{sec:experiment} 
The experimental setup in this study consists of a modified version of the ESI introduced in \cite{electrospraying_chen_1995}, the "Uppsala"-injector \cite{hantke2018rayleigh} with a two-skimmer box setup, an optical scattering setup to detect the nanoparticles in the main chamber \cite{Yenupuri2023bionanobeams} and a residual gas analyzer (RGA) (Extorr Inc., XT100M) to analyze the gas composition inside the chamber.

\subsection{Modified ESI source: He-ESI design}
\begin{figure}[H]
  \centering
  \includegraphics[scale=0.9]{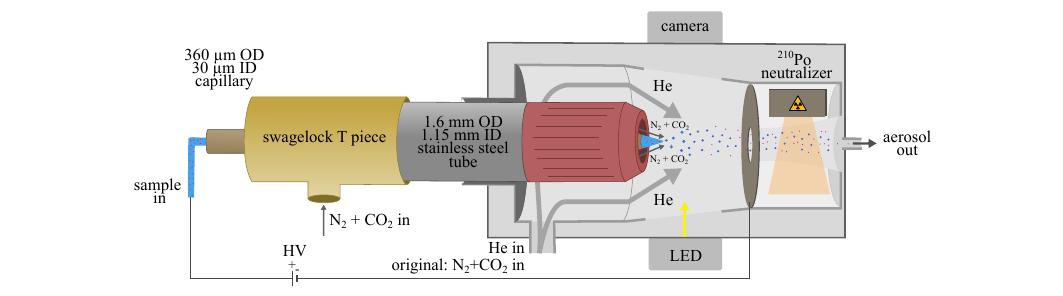}
    \caption{Schematic of the He-ESI. The modification of the ESI to operate with He includes a Swagelok T-piece, a stainless steel tube and the 3D-printed Uppsala nozzle. Liquid sample flows through the capillary and a stable Taylor cone is formed by applying a high voltage. In between the capillary and the inside of the nozzle a N$_2$ and CO$_2$ environment is formed with a combined flow rate of around 50 mL/min. Helium is introduced through the original gas inlet, surrounding the nozzle within the electrospray head and facilitating the flow of particles. The highly charged droplets pass through a Po-210 neutralizer. Then, the neutralized aerosol is exiting the electrospray head.}
    \label{fig:He-ESI}
\end{figure}

The modified ESI setup is shown in Figure \ref{fig:He-ESI}. It includes a 3D printed nozzle (Uppsala nozzle) measuring 4.45 x 1.56 x 1.56 mm$^3$ printed via two-photon polymerization in a liquid resin (UpPhoto) within 35 minutes using the NanoOne 3D printing system (UpNano). 
After printing, the nozzle was glued to a stainless-steel tube with an inner diameter (ID) of 1.15 mm using a standard two-component epoxy glue (Loctite power epoxy) and connected to the N$_2$ and CO$_2$ gas mixture line. To reduce the background scattering in SPI experiments, we replaced most of the N$_2$ and CO$_2$ used for particle transport with He. The gas inlet previously used for the N$_2$ and CO$_2$ gas mixture was used as the He inlet, as shown in Figure \ref{fig:He-ESI}. 
The Uppsala nozzle is designed to hold the silica fused capillary of \SI{360}{\um} outer diameter (OD) in the center of the nozzle as shown in Figure \ref{fig:He-ESI}. We reduced the consumption of N$_2$ and CO$_2$ by placing a 3D-printed structure around the capillary generating an N$_2$ and CO$_2$ atmosphere between the capillary and the nozzle and filling the rest of the ESI head with He.

An alternative 3D-printed nozzle design, referred to as the EuXFEL nozzle, follows the same principle of gas replacement but does not require the use of a fused silica capillary inside. Instead, it is entirely printed using the Nanoscribe Photonic Professional GT with IP-S photoresist. This design incorporated two capillary inlets: one with an ID of \SI{40}{\um} for the sample and another with an ID of 180 µm for protective gases (N$_2$ and CO$_2$). The dimensions of the EuXFEL nozzle are $1.4 \times 0.5 \times 1.2$ mm. Details on the EuXFEL nozzle can be found in the supplementary materials. Furthermore, the CAD models for both the Uppsala and EuXFEL nozzles are freely accessible and can be downloaded from our GitHub repository at (\url{https://github.com/ytejvarma/Helium-nozzle}) and (\url{https://github.com/safirafie/ESDesign}) respectively.

\subsection{Simulations of Gas Flow Around the Taylor Cone}
To protect the Taylor cone from Corona discharge and to minimize heavier gases, it is important to understand the behaviour of gases surrounding the Taylor cone within the He-ESI system. Therefore, we performed simulations using COMSOL Multiphysics, a finite element analysis software \cite{comsol2022}. 
We used the laminar flow interface and the transport of concentrated species interface and coupled these interfaces together through a multi-physics interface. The laminar flow interface allowed us to model the gas flow dynamics by computing the velocity and pressure fields of the gases. Concurrently, the transport of concentrated species interface was used to study gaseous mixtures by solving for the mass fractions of all participating species.

To monitor the risk of corona discharge around the Taylor cone, we calculated the fractional concentration of each gas, denoted as $x_{i}$ and defined as:
$$
x_{i}  =  \frac{c_{i}}{c_{i}    + c_{j}   + c_{k} }
$$
where $c_{i}$ symbolizes the molar concentration of the gas for which we are determining its fractional concentration, $x_{i}$. The $c_{j}$ and $c_{k}$ denote the molar concentrations of the remaining two gases in the mixture.

\begin{figure}[H]
  \centering
  \sidesubfloat[]{\includegraphics[scale=0.23]{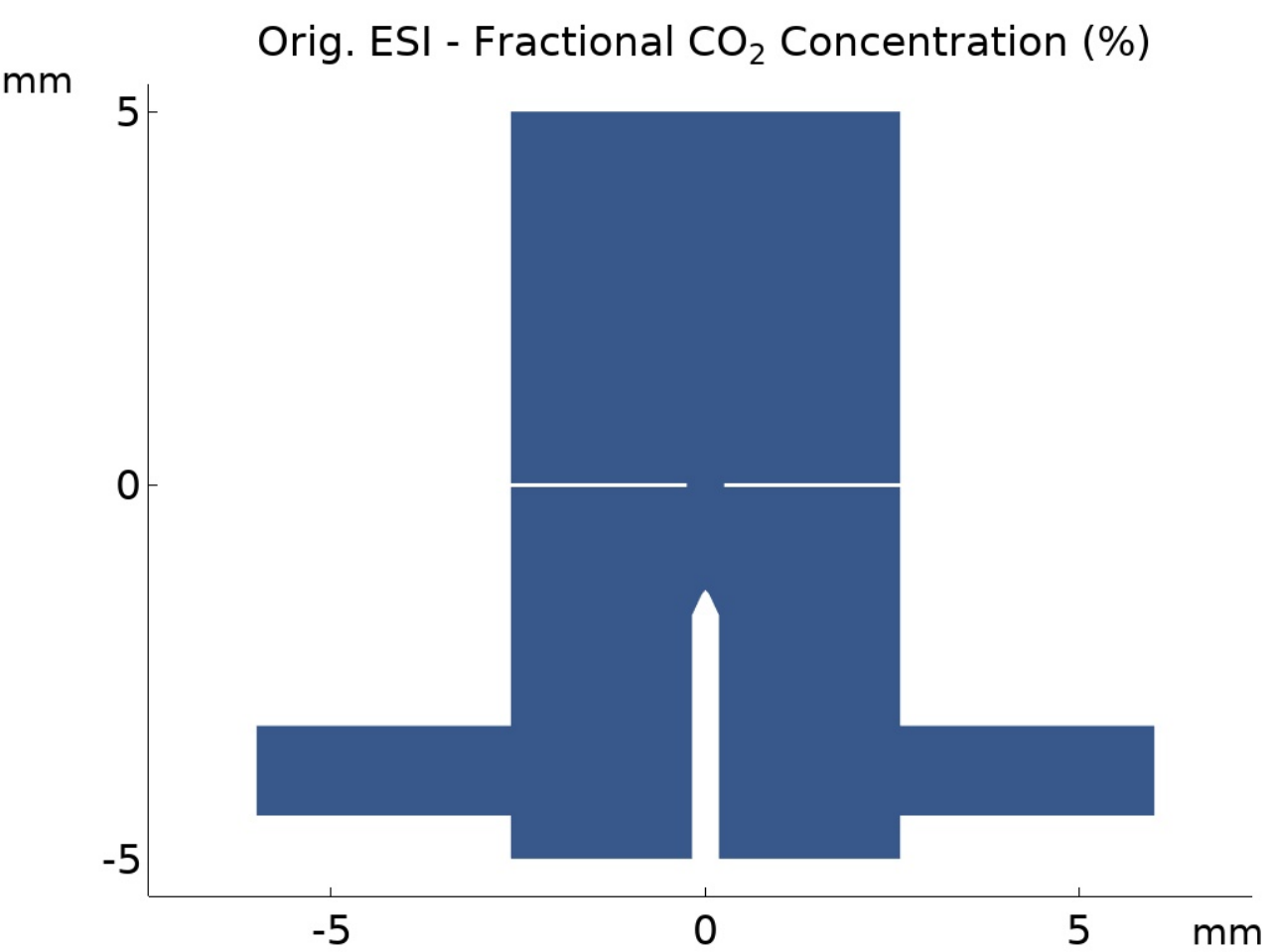}\label{fig:he_esi_He_a}}
  \sidesubfloat[]{\includegraphics[scale=0.23]{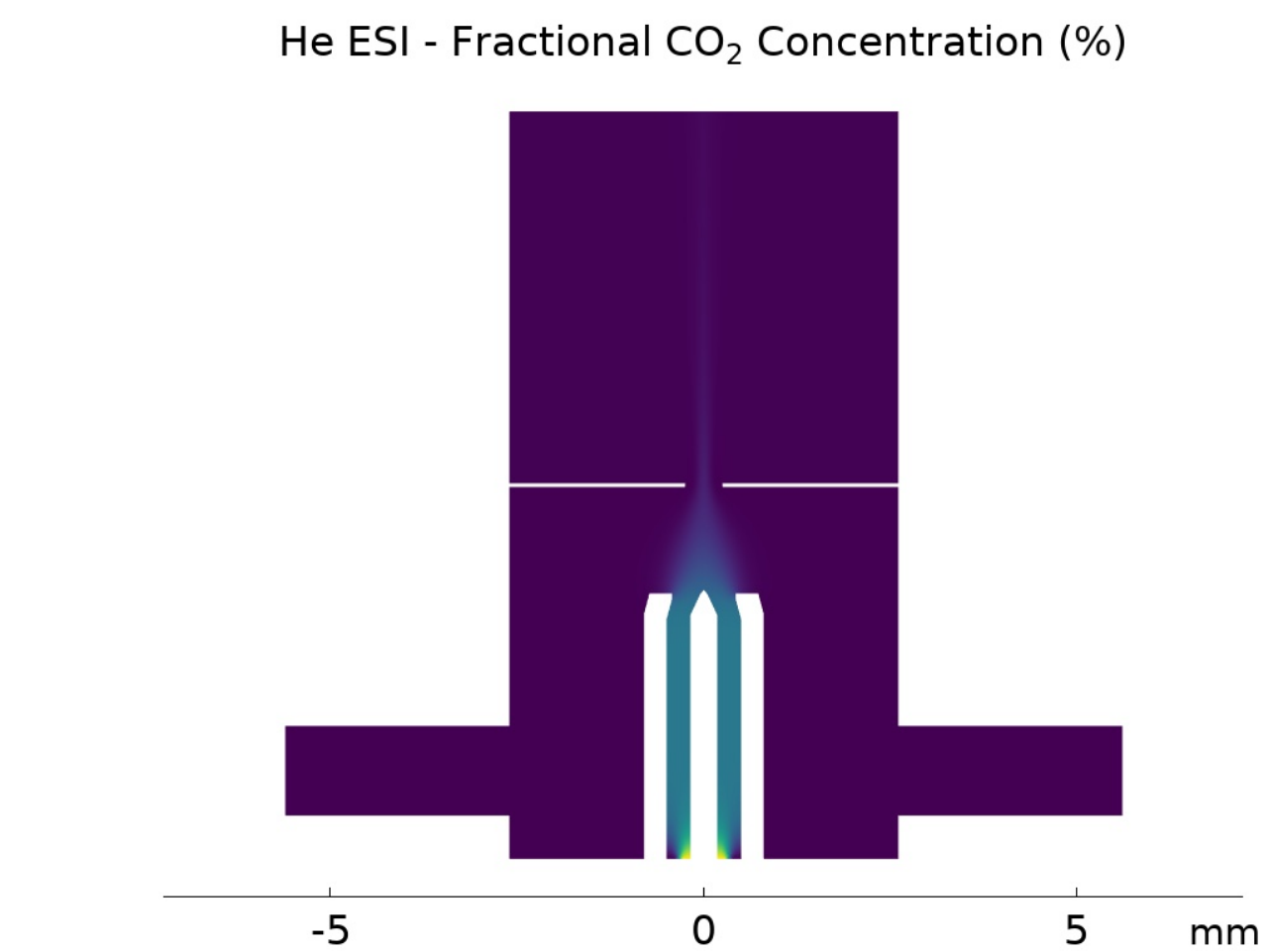}\label{fig:he_esi_He_b}}
  \sidesubfloat[]{\includegraphics[scale=0.23]{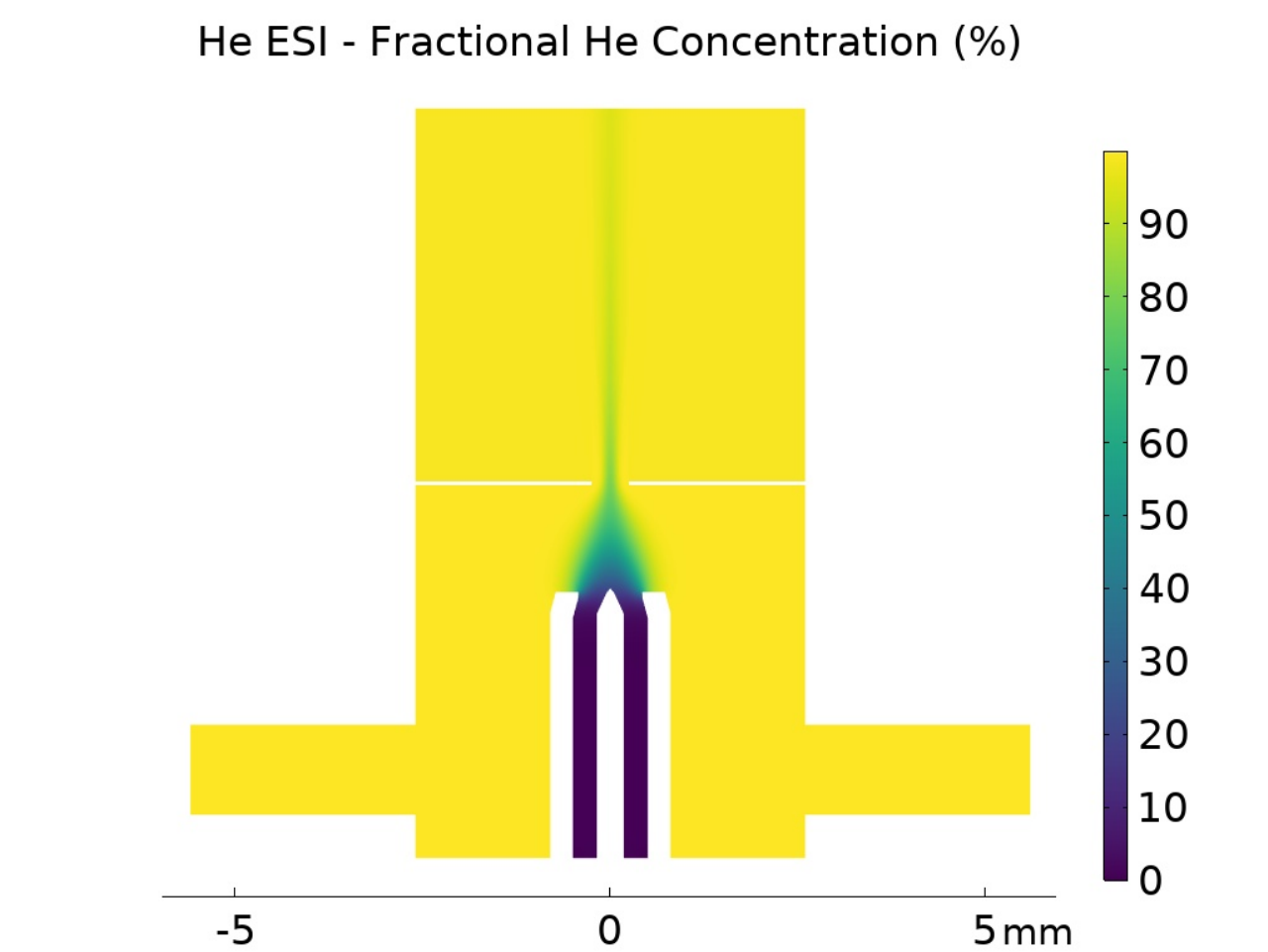}\label{fig:he_esi_He_c}}
  \hspace{0.1in}

  \sidesubfloat[]{\includegraphics[scale=0.3]{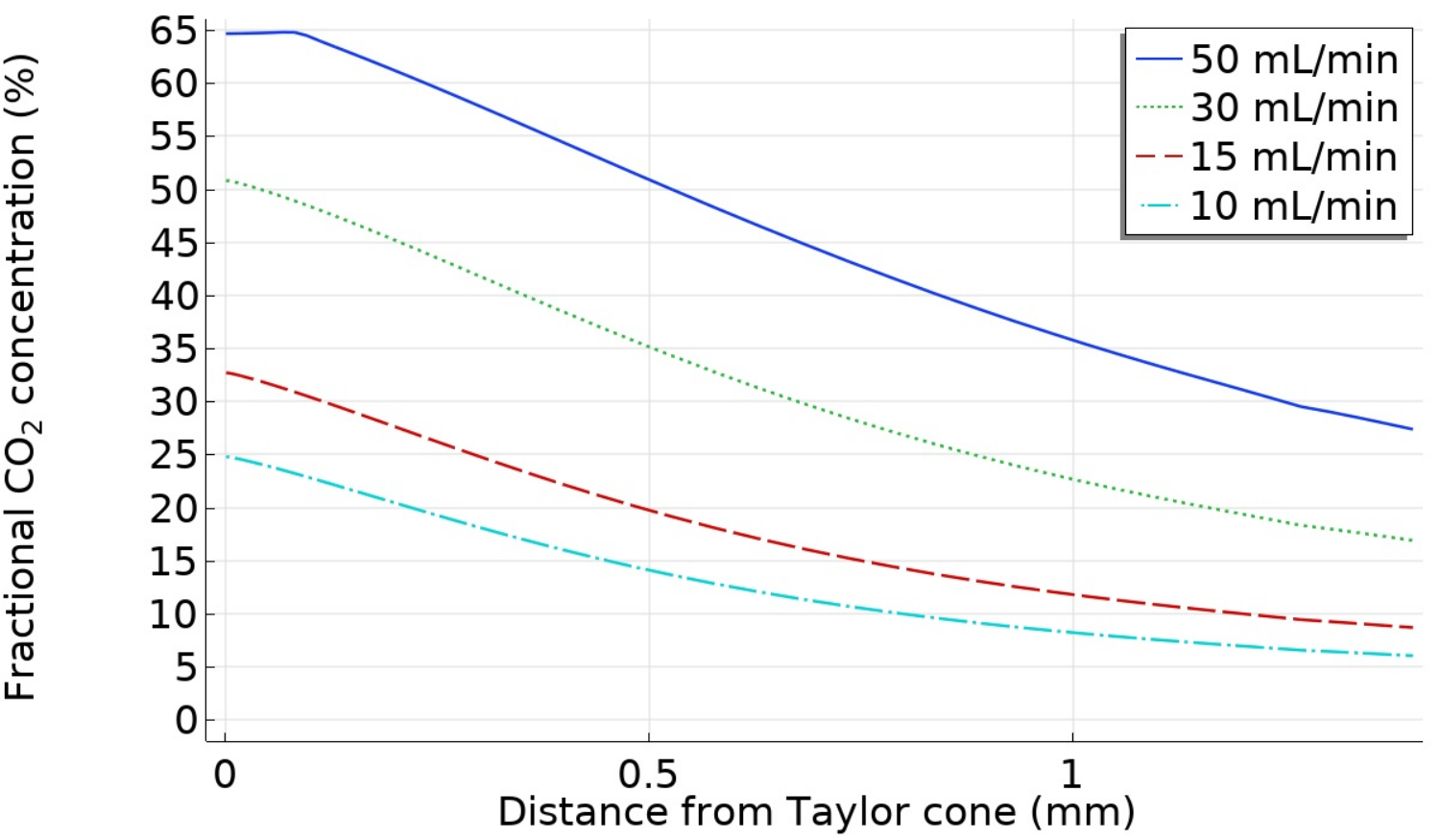}\label{fig:he_esi_He_d}}
  \hspace{0.02in}
  \sidesubfloat[]{\includegraphics[scale=0.3]{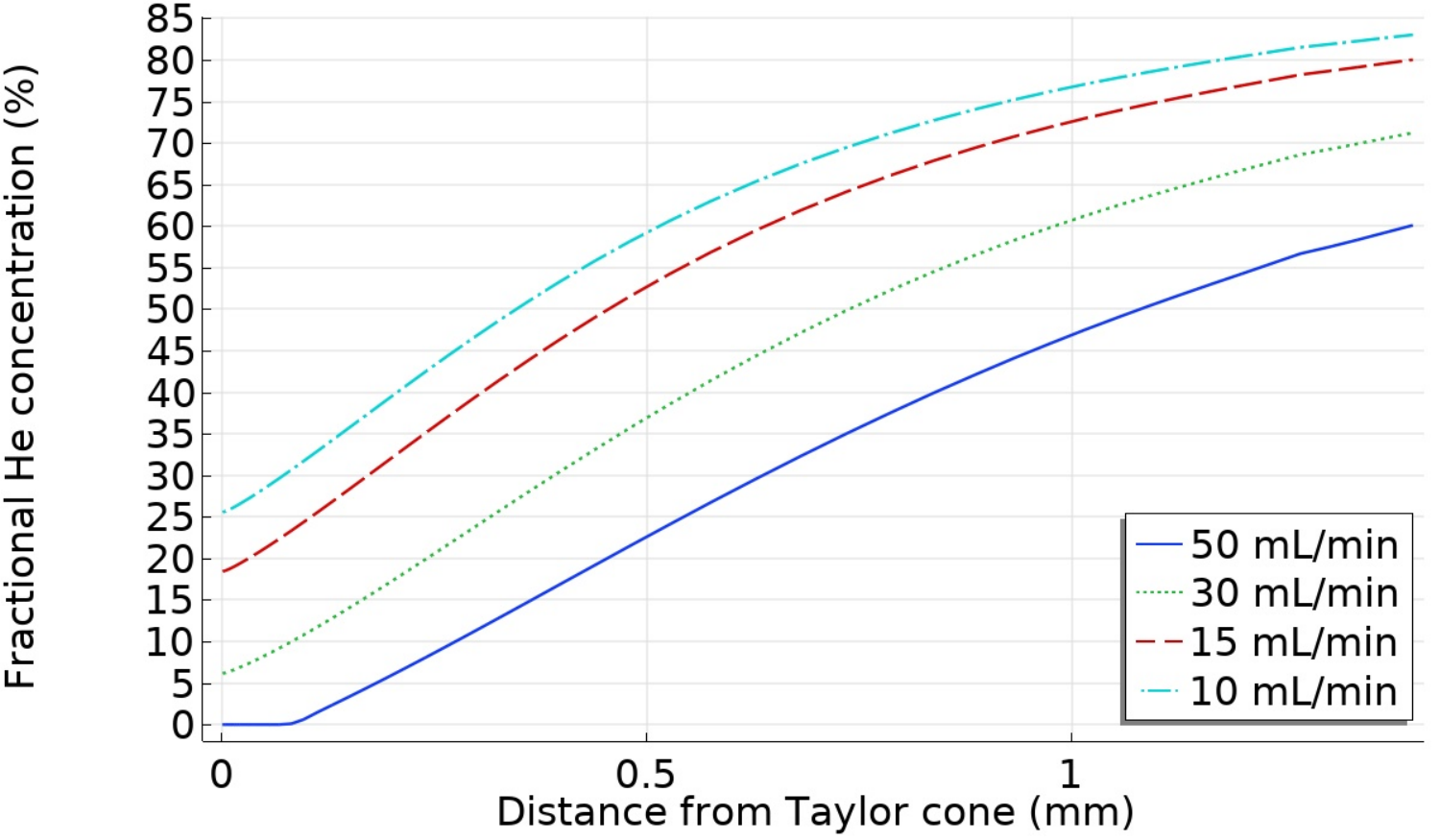}\label{fig:he_esi_He_e}}
    \caption{The fractional concentration of various gases in the vicinity of the Taylor cone, highlighting the effectiveness of gas shielding against corona discharge. a) The fractional concentration of CO$_2$ in the original ESI system. b) and c) The fractional concentration of CO$_2$ and He respectively in the He-ESI system. d) and e) The fractional concentration of CO$_2$ and He respectively in the He-ESI system under varying CO$_2$ flow rates 
    while maintaining a constant He flow rate of 1.2 L/min and N$_2$ flow rate of \SI{20}{\milli\liter\per\min}.}
    \label{fig:he_esi_He}
\end{figure}

We compared the gas distribution between the original ESI and the He-ESI system, as displayed in Figures \ref{fig:he_esi_He_a}, \ref{fig:he_esi_He_b}, and \ref{fig:he_esi_He_c}. In the original ESI system, a mixture of two gases was used. N$_2$ was utilized as the carrier gas at a flow rate of \SI{1}{\liter\per\min}, and CO$_2$ was employed as a protective gas, to shield the Taylor cone from corona discharge, with a flow rate of \SI{150}{\milli\liter\per\min}. The He-ESI system instead uses a mixture of three gases. He serves the role of the carrier gas with a flow rate of \SI{1.2}{\liter\per\min}, while N$_2$ and CO$_2$, with flow rates of \SI{20}{\milli\liter\per\min} and \SI{15}{\milli\liter\per\min} respectively, functioned as protective gases. The simulation results illustrate how in the He-ESI system, the Taylor cone is effectively enveloped by CO$_2$, preventing corona discharge. 

To study the influence of various CO$_2$ flow rates on the gas distribution around the Taylor cone in the He-ESI setup, we performed simulations at CO$_2$ flow rates of 10, 15, 30, and \SI{50}{\milli\liter\per\min}, as depicted in Figures \ref{fig:he_esi_He_d} and \ref{fig:he_esi_He_e}. The He and N$_2$ flow rates were kept constant at \SI{1.2}{\liter\per\min} and \SI{20}{\milli\liter\per\min}, respectively. 
These simulations help estimate the minimal fractional concentration of gases necessary to sustain a stable Taylor cone, thus minimizing the potential for corona discharge. This provides important insights into the interactions and flow dynamics of the gases around the Taylor cone, which can further help us optimize the design and operating conditions of the electrospray system.

\subsection{Operating Conditions for the He-ESI}
The He-ESI with the Uppsala nozzle is stable under the following conditions: the tip of the angled capillary (conically ground at an angle of 30\textsuperscript{o}) must be kept at the edge or slightly inside the nozzle, which is placed at a distance of 1.5 - 1.8 mm away from the grounded orifice of 0.5 mm diameter. The liquid sample flow rates must be 100 - 200 nL/min, the He flow rate in the ESI head should be 1.2-1.4 L/min, the N$_2$ flow rate 0.03-0.035 L/min, the CO$_2$ flow rate 0.015-0.02 L/min and the voltage between 2.2 - 2.6 kV. The stability of the He-ESI was monitored by measuring the current (typically around 200 - 300 nA) and visually with a camera pointing at the Taylor cone. Under these conditions, the Taylor cone ejects charged droplets. These charged droplets pass through a Po-210 neutralizer, which neutralizes the charge on the particles and the neutralized particles then pass through a conductive tubing to the inlet of the experimental setup. 
Detailed information regarding the operating conditions using the EuXFEL nozzle can be found in the supplementary materials.

\subsection{Injector Setup: Operation using He-ESI}
The He-ESI is coupled to the injector setup \cite{electrospray_bielecki_2019}. An extra helium inlet was added to the injector setup before the first skimmer stage to avoid the suction of the gas in the aerosolization and neutralization chamber of the ES due to the pumping in the skimmer stages and to protect the Taylor cone. 
Typically, 2.5-\SI{3}{\liter\per\min} He is added at the aerosol inlet.  
In total, \SI{4.2}{\liter\per\min} He is required in the setup. The excess gas is skimmed away using scroll pumps at the two nozzle-skimmer stages. The particles enter the aerodynamic lens with a pressure of 1-1.2 mbar 
and exit the lens through a 1.5 mm aperture into the interaction region in the experimental chamber, which is kept at 10$^{-5}$ mbar.

\subsection{Gas Reduction in the Interaction Chamber}
\begin{figure}[H]
  \centering
  \includegraphics[scale=1]{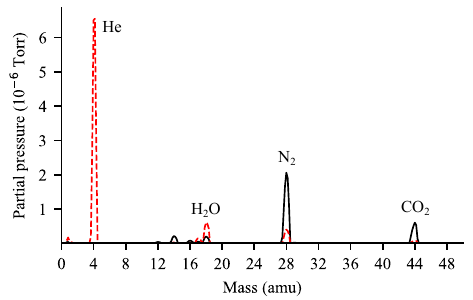}
    \caption{Residual gas analysis spectrum inside the interaction chamber. Measured for the He-ESI (dashed red) with flow rates of 4.2 L/min He, 0.03 L/min N2, 0.015 L/min CO2 and the original ESI (solid black) with flow rates of 1 L/min N2 and 0.2 L/min CO2.}
    \label{fig:He-ESI_RGA}
\end{figure}

We used an RGA, mounted 25 cm away from the interaction region, to determine the composition of the gas in the interaction chamber. RGA spectra while using both types of ESI are shown in Figure \ref{fig:He-ESI_RGA}.
For the He-ESI (dashed red line), the largest contribution is He at 4 atomic mass units (amu) with a partial pressure, measured from the peak area, of $1.9\times 10^{-5}$ Torr, while N$_2$ and CO$_2$, shown in the spectrum at 28 and 44 amu, have partial pressures of $1.6\times 10^{-6}$ Torr and $3.1\times 10^{-7}$ Torr respectively. There's a further peak at 18 amu due to water contamination.

The relative composition of the input gases to the He-ESI is 1.22 \% N$_2$ and 0.97 \% CO$_2$, compared to 8 \% N$_2$ and 1.5 \% CO$_2$ measured in the interaction chamber. This discrepancy may be explained by the different pumping efficiency for He, N$_2$ and CO$_2$ based on Graham's law, which states that the rate of diffusion or effusion of gas is inversely proportional to its molecular weight. This implies, that N$_2$ and CO$_2$ diffuse much slower than He when passing through the nozzle in the two skimmer stages leading to He being skimmed away first and more efficiently.

The RGA spectrum of the original ESI, shown in black, shows much larger N$_2$ and CO$_2$ peaks, with partial pressures of $8.7\times 10^{-6}$ and $2.4\times 10^{-6}$ Torr respectively. 

While the gases in the interaction chamber will scatter both elastically and inelastically, the inelastically scattered photons can be filtered due to their different energy. But the elastically scattered ones are indistinguishable from those scattered by the sample and are the main contributors to background noise in SPI \cite{ekeberg2022observation}.
For the resolutions relevant to SPI, each gas molecule is well approximated as a point scatterer and the total scattering is then proportional to the square of the number of electrons. We can then estimate the elastic scattering from the gas as the weighted sum of the contributions of the different gas species, and with it calculate the expected elastic scattering by the gas when using the He-ESI relative to the original ESI ($I_{rel}$),


$$
I_{rel}  =  \frac{ p^{\text{new}}_{N_2} Z_{N_2}^2 + p^{\text{new}}_{CO_2} Z_{CO_2}^2 + p^{\text{new}}_{He} Z_{He}^2}{p^{\text{old}}_{N_2} Z_{N_2}^2 + p^{\text{old}}_{CO_2} Z_{CO_2}^2},
$$

where $p^\text{new}$ are the partial pressures of the He-ESI setup, $p^\text{old}$ of the original ESI and $Z$ is the total number of electrons of each gas molecule. Using this equation with the partial pressures measured above we obtain an $I_{rel}$ of 0.188 or an expected reduction of scattering intensity by $\approx 81$ \%.

\subsection{Sample delivery performance with the He-ESI}
To show that the He-ESI is working stable and generates aerosolized particles suitable for SPI experiments, we coupled the He-ESI to the “Uppsala”-injector. To detect the flow of particles into the interaction region we used a Rayleigh-scattering microscopy setup \cite{Yenupuri2023bionanobeams} and recorded particle intensities and beam evolution curves for 20-80 nm polystyrene spheres (PS) in a 20 mM ammonium acetate (AmAc) buffer solution. The beam-evolution curve is shown in the supplementary information Figure S2. At a given injector pressure, the particle-beam focus position moves away from the injector exit with increasing particle size. A similar behaviour has been observed in a previous study using the same injector and GDVN aerosolization, i.e. focusing with He, for PS with diameters larger than 40 nm \cite{hantke2018rayleigh}. For larger particles, the gas density in the focus is lower. In addition, the opening angle of the particle beam decreases with increasing particle size. The particle-beam parameters are summarized in Tab. S1. The data collection and analysis are discussed in a \cite{Yenupuri2023bionanobeams}. 
Next, we characterized the particle-beam density at the particle beam focus for different sizes of PS and Bacteriophage MS2 (MS2) and compared it to particle-beam density measurements using the original ESI for aerosolization, i.e. using N$_2$ as the main carrier gas. As a proxy for the particle-beam number density, we show the number of particles collected in 1000 frames. Each frame contains one laser pulse for particle detection. Table \ref{tab:long} shows the measured mean number of particle hits per 1000 frames in the particle-beam focus. For all particle sizes, the measured number of particles is higher using the He-ESI compared to the original ESI. While the improvement of the measured particle numbers is different for the used sizes, the highest improvement in particle throughput, by a factor of $\approx$ 11, is observed in the bioparticle MS2. 

\begin{table}[ht]
\centering
\rowcolors{2}{gray!10}{white}
\caption{Comparison between He-ESI and original ESI of the mean number of particles per 1000 frames as a function of the sample diameter.}
\label{tab:long}
\begin{tabularx}{0.6\textwidth}{ X S S}
\hline
\rowcolor{white}
{} & \multicolumn{2}{c}{Particles per 1000 frames} \\
{Sample/DMA size~(nm)}  & {He-ESI}  & {Original-ESI} \\ 
\hline
Bacteriophage MS2/ 25.9 & {460 } & 42\\
20 nm PS/ 18.9 & {1010} & 271\\ 
30 nm PS/ 28.9 & {2546}  & 517\\ 
40 nm PS/ 42.9 & {2264} & 874\\
50 nm PS/ 59.4 & {1553} & 939\\ 
70 nm PS/ 76.4 & {1118} & 527\\ 
80 nm PS/ 88.2 & {1150} & 300\\
\hline
\end{tabularx}
\newline

\end{table}

\subsubsection{Exploration of Various Ionization Techniques in He-ESI}
A comparative study was conducted to analyze the transmission efficiency between two different ionization techniques used in He-ESI: a Polonium (Po-210) source and an ultraviolet (UV) ionizer. The target sample utilized for this experiment was a silver cube suspended in ethanol. In both techniques, the gas flow rates were maintained at 1 L/min for He and 30 mL/min for CO$_2$. Particle detection was carried out in the interaction chamber using Rayleigh scattering \cite{Yenupuri2023bionanobeams}. The results from the number of particles detected showed that the Polonium source delivered approximately 5\% more particles than the UV ionizer to the interaction chamber.

Nonetheless, given that UV light is more efficient at ionizing N$_2$ gas \cite{zinedine2023}, we extended our experiment by adding 30 mL/min of N$_2$ through the He inlet. This introduction of N$_2$ enhanced the transmission efficiency of the UV ionizer setup and outperformed the Polonium source setup by delivering approximately 30\% more particles. This enhancement can be attributed to the improved neutralization of the particles, facilitated by the UV ionizer's more effective in ionizing N$_2$. These findings suggest that the inclusion of N$_2$ gas in the UV ionizer setup could be a potential strategy to enhance transmission efficiency in He-ESI. It is important to highlight, though, that the advantages gained from incorporating N$_2$ need to be balanced against its potential contribution to background noise.

\section{Discussion and Outlook}
Within this paper, we presented improvements in the sample aerosolization process by developing a He-ESI to reduce the background scattering due to gases in SPI experiments. We used 3D printed nozzles to reduce the amount of N$_2$ and CO$_2$ and kept modifications of the previously used ESI setup to a minimum. With the He-ESI, the main particle transport gas into the interaction chamber is He. 
In the interaction chamber and based on RGA measurements using the He-ESI with the Uppsala-nozzle, the amount of N$_2$ was reduced by 82 \% and for CO$_2$ by 87.7 \%. While the large reduction of the heavy gasses in the initial gas mixture could not be observed to the same extent in the interaction region, presumably due to different pumping efficiencies, an optimization of the skimmer assembly may improve the ratio in the interaction region further. Nonetheless, assuming the ratio of the gasses measured in the RGA translates into the ratio of contribution to background scattering, we reduced the gas background scattering off the gas by 81 \%.

Additionally, through simulations conducted using COMSOL Multiphysics, our study has deepened the understanding of gas flow dynamics around the Taylor cone in a He-ESI system. This allowed us to model the behaviour of different gas mixtures, examining their respective impacts on protecting the Taylor cone from corona discharge. Given our optimal operational conditions with a water-based buffer our simulations suggest that to maintain a stable Taylor cone, the He percentage should not exceed 20\% at the cone's tip. Further computational analysis of the gas distribution and breakdown voltage can aid in determining the minimum fractional concentration necessary to maintain a stable cone before corona discharge occurs.

Our modification of the ESI not only demonstrates a decreased use of heavy gasses for sample injection but also an increased throughput of particles into the interaction region. The highest increase in transmission of particles was observed while injecting small bioparticles: approximately by a factor of 11 for MS2 particles. Whereas, while delivering PS into the interaction region, we measure an increase in the transmission of particles by a factor of 2 to 5 depending on the particle size.

To further enhance particle transmission, we conducted a comparative analysis of the transmission efficiency between Po-210 sources and UV ionizer techniques within He-ESI systems. Our results demonstrated that by adding 30 mL/min of N$_2$ gas along with He at the He inlet, the UV ionizer's performance was enhanced, surpassing the Po-210 source by approximately 30\%. For a more comprehensive understanding of their impact on transmission efficiency, future studies could investigate the simultaneous utilization of both the Po-210 source and the UV ionizer in He-ESI systems.

Although this is not the first adaptation of ES injection for X-ray diffractive imaging, the presented modification is a much-required leap towards single protein imaging by aiming at lower background scattering from the injection gases, allowing us to recognize lower scattering signals from the sample in the diffraction data. We expect the He-ESI to improve the quality of collected data and provide better experimental conditions for X-ray imaging of small nanoparticles not only due to the lowered background but also because of a higher particle transmission through the injector. Together, higher quality and quantity of diffraction patterns can be collected in the future using a He-ESI for sample aerosolization.

\bibliographystyle{naturemag}
\bibliography{references}

\end{document}


\maketitle
\begin{tabular}{@{}l@{}}
\(^{\ddag}\)These authors contributed equally to this work.\\ 
\(^{*}\) Correspondence e-mail: johan.bielecki@xfel.eu, filipe.maia@icm.uu.se
\end{tabular}
\vspace{0.3in}

\section{METHODS AND RESULTS}

\subsection{He-ESI Design: the EuXFEL Nozzle} 
\begin{figure}[H]
  \centering
  \includegraphics[scale=0.13]{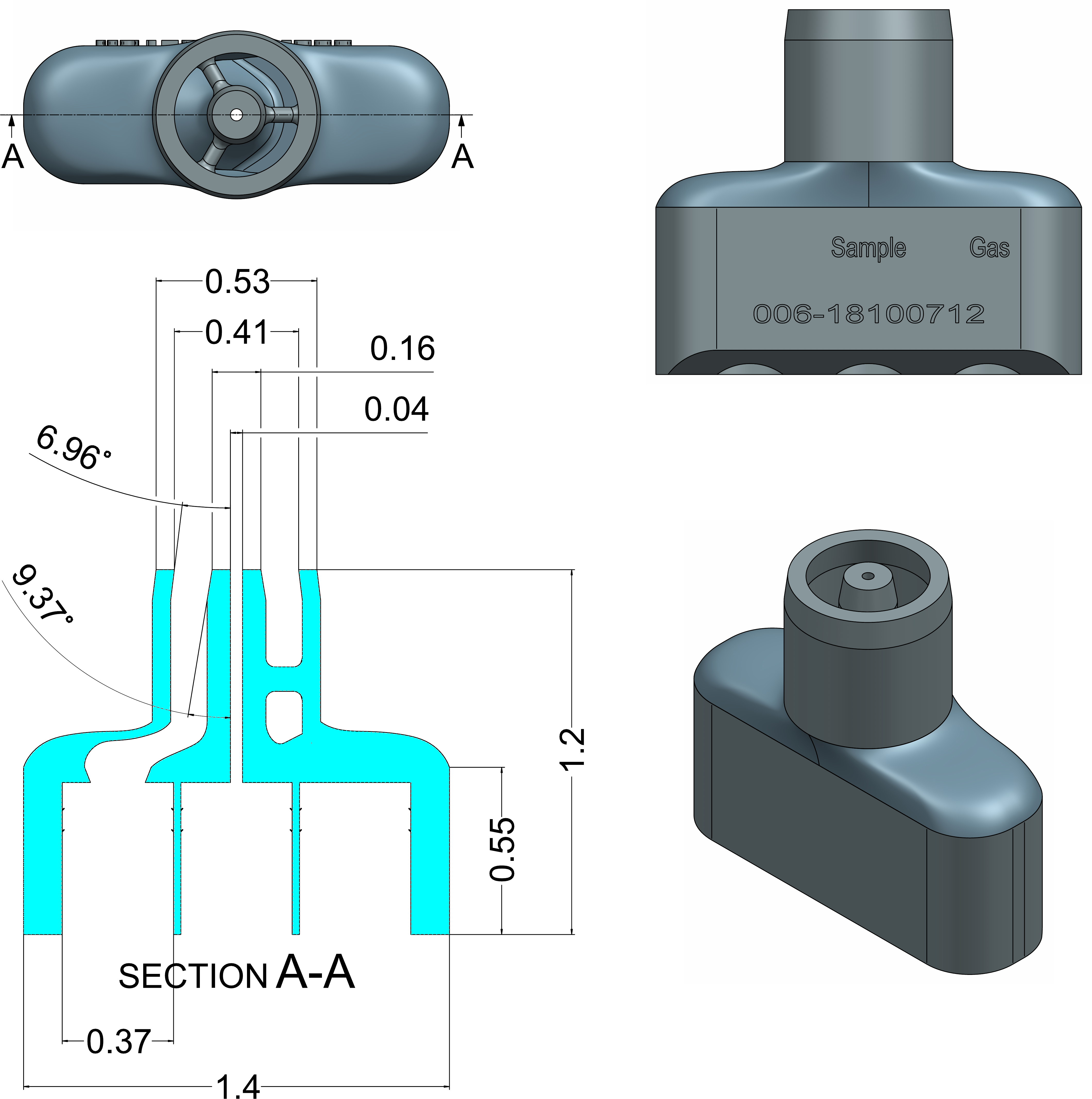}
    \caption{Schematic drawings of the EuXFEL Nozzle illustrate the dimensions and depict the inlets and outlets of the nozzle.}
    \label{fig:006}
\end{figure}

The EuXFEL nozzle was engineered using Siemens' NX software and was designed with three capillary inlets suitable for \SI{360}{\um} outer diameter (OD) fused silica capillaries for fluid feed, along with two outlets as shown in Figure \ref{fig:006}. The inlet ports comprised one for a sample with an inner diameter (ID) of \SI{40}{\um}, another for gas with an ID of \SI{180}{\um}, and a dummy one which aids in centering the sample capillary. The outlet ports included one designated for the sample, with an ID of \SI{40}{\um} and an angle of approximately \SI{10}{\degree}, and another, concentric with the first, designated for gas, with an ID of \SI{410}{\um} and an angle of approximately \SI{7}{\degree}.

The nozzle design was outputted in STL formats. The conversion of these STL-based 3D designs into print-job instructions, or GWL, was executed using Nanoscribe's DeScribe software. For better structural stability of the fabricated devices, we adopted a solid volume printing strategy with slicing of \SI{1}{\um} and hatching of \SI{0.5}{\um}. The devices were then printed using the Nanoscribe Photonic Professional GT with IP-S photoresist as the printing material. The process utilized a 25x objective lens from Zeiss, full laser power, and a printing velocity of \SI{100,000}{\um\per\second}. Under these conditions, the printing duration for a single device was approximately one hour.

Following the printing process, the glass slide with the cured photoresist was submerged in a beaker of propylene glycol methyl ether acetate (PGMEA) for one or two days to dissolve any remaining uncured parts, a process known as development. Post-development, the devices were transferred to a beaker of isopropanol for about 30 min, then relocated to another beaker filled with fresh isopropanol. Finally, the devices were left on a cleanroom cloth to dry under ambient conditions.

The nozzles were assembled on clean polydimethylsiloxane (PDMS) sheet, with the process monitored under an optical microscope. To secure the devices, an additional piece of PDMS was applied over them. Following this, three fused silica capillaries, each with an OD of \SI{360}{\um}, were inserted into their designated fluid inlets on the nozzle and secured with a 5-minute epoxy glue from Devcom. These capillaries were then guided through hollow stainless-steel tubing with an OD of 1/16 inch (IDEX U-145 with an ID of 0.046 inches) and glued between the nozzle material and steel.

\subsection{Operating Conditions for the EuXFEL Nozzle} 
The operating stability of the He-ESI system is influenced by factors such as the buffer type, the buffer conductivity, and the geometry of the aerosolization chamber. To minimize the presence of heavier gases and maintain a stable Taylor cone, the operating conditions were carefully optimized. Experiments were conducted using the EuXFEL nozzle with two different buffers: water (with conductivities ranging from 900 to 1600 $\mu$S/cm) and ethanol (with conductivities ranging from 800 to 1300 $\mu$S/cm). With the water buffer, we used a He flow rate of $1-1.5$ L/min, a N$_2$ flow rate of $20-30$ mL/min, and a CO$_2$ flow rate of $15-25$ mL/min. With the ethanol buffer, the He flow rate was adjusted to $1-1.6$ L/min, while the CO$_2$ flow rate was set at $10-20$ mL/min, without any N$_2$ flow. The nozzle was tested with two different ionizers: a Po-210 source and a UV ionizer, before transporting the particles to the Uppsala injector.

\subsection{PS particle-beam parameters}
\begin{figure}[ht]
  \centering
  \includegraphics[scale=1]{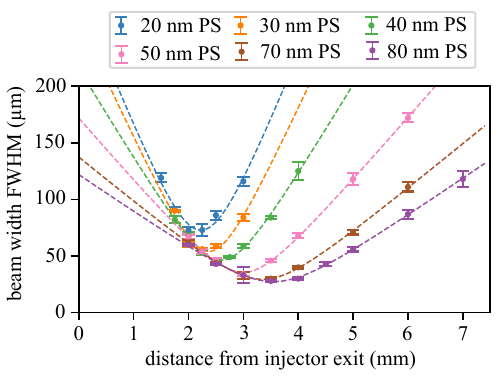}
    \caption{Particle-beam evolution curves for 20 - 80 nm PS at 1.0 mbar injector pressure using the He-ESI with Uppsala nozzle for aerosolization. The main focussing gas is He.} 
    \label{fig:beam-evolution}
\end{figure}

The particle-beam width depending on the distance from the injector exit was measured for different sizes of PS. A Gaussian beam evolution fit was used to determine the focus width and the focus position. The particle-beam evolution curves are shown in Figure \ref{fig:beam-evolution} and the focus values are summarized in Table \ref{tab:supplementalParameters}. A clear shift of the particle-beam focus towards the injector exit with decreasing particle size is observed and the particle-beam focus width increases as the particle size decreases. 

\begin{table}[ht]
\centering
\caption{Experimental particle-beam parameters (focus position and width) for different sizes of PS. The particles were aerosolized using the He-ESI and the injector pressure was kept constant at 1.1 mbar. }
\label{tab:supplementalParameters}
\begin{tabular}{|c|c|c|}
\hline
{Sample/DMA size (nm)}  & {focus position (mm)}  & {focus width FWHM (µm)}\\ 
\hline
20 nm PS/ 18.9 & {2.23} & 73\\ 
30 nm PS/ 28.9 & {2.37}  & 54\\ 
40 nm PS/ 42.9 & {2.59} & 46\\
50 nm PS/ 59.4 & {2.99} & 36\\ 
70 nm PS/ 76.4 & {3.35} & 29\\ 
80 nm PS/ 88.2 & {3.55} & 27\\
\hline
\end{tabular}
\end{table}